\documentclass[twocolumn,showpacs,preprintnumbers,amsmath,amssymb, superscriptaddress, prb]{revtex4-1}

\usepackage{graphicx}
\usepackage{dcolumn}
\usepackage{bm}
\newcommand{\FeTe} {{\mbox{FeSe${}_{0.42}$Te${}_{0.58}$ }}} 
\newcommand{\Te} {{\mbox{${}^{125}$Te }}} 
\newcommand{\Se} {{\mbox{${}^{77}$Se }}}

\usepackage{color}
\definecolor{red}{rgb}{1,0,0}

\definecolor{blue}{rgb}{0,0,1}

\definecolor{green}{rgb}{0,1,0}

\begin{document}
\preprint{APS}

\title{Two-electronic component behavior in the multiband FeSe${}_{0.42}$Te${}_{0.58}$ superconductor}
\author{D. Ar\v{c}on}
\email{denis.arcon@ijs.si}
\affiliation{Institute "Jozef Stefan", Jamova 39, 1000 Ljubljana, Slovenia}
\affiliation{Faculty of Mathematics and Physics, University of Ljubljana, Jadranska 19, 1000 Ljubljana, Slovenia}
\author{P. Jegli\v{c}}
\affiliation{Institute "Jozef Stefan", Jamova 39, 1000 Ljubljana, Slovenia}
\author{A. Zorko}
\affiliation{Institute "Jozef Stefan", Jamova 39, 1000 Ljubljana, Slovenia}
\author{A. Poto\v{c}nik}
\affiliation{Institute "Jozef Stefan", Jamova 39, 1000 Ljubljana, Slovenia}
\author{A. Y. Ganin}
\affiliation{Department of Chemistry, University of Liverpool, Liverpool L69 7ZD, UK}
\author{Y. Takabayashi}
\affiliation{Department of Chemistry, Durham University, Durham DH1 3LE, UK}
\author{M. J. Rosseinsky}
\affiliation{Department of Chemistry, University of Liverpool, Liverpool L69 7ZD, UK}
\author{K. Prassides}
\affiliation{Department of Chemistry, Durham University, Durham DH1 3LE, UK}
\date{\today}

\begin{abstract}
We report X-band EPR and  \Te  and \Se NMR measurements on single-crystalline superconducting \FeTe ($T_c$ = 11.5(1) K). The  data provide evidence for
the coexistence of intrinsic localized and itinerant electronic states. In the normal state, localized moments couple to itinerant electrons in the Fe(Se,Te) layers and affect the local spin susceptibility and spin fluctuations.  Below $T_c$,  spin fluctuations become rapidly suppressed and an unconventional  superconducting state emerges in which $1/T_1$  is reduced at a much faster rate than expected for conventional $s$- or $s_\pm$-wave symmetry. We suggest that the  localized states arise from the strong electronic  correlations within one of the Fe-derived bands. The multiband electronic structure together with the  electronic correlations thus determine the normal and superconducting states of the FeSe$_{1-x}$Te$_x$ family, which appears much closer to other high-$T_c$ superconductors than previously anticipated.  
\end{abstract}

\maketitle

The competition between magnetism and superconductivity (SC) is a common theme in all high-$T_c$ superconductors. In cuprates \cite{cup} and fullerides,\cite{sci,nat} the parent compounds are antiferromagnetic (AF) Mott insulators and SC appears after  doping with charge carriers or upon applying high pressure. Strong electron correlations are the driving force in these transitions. On the other hand, in iron-based superconductors their role in SC is less clear. They are believed to be  weaker \cite{yang}  because the ground state of the parent compounds is  metallic and  a low-temperature spin density wave (SDW) instability is induced by the Fermi surface nesting.   
 Iron-based superconductors have a complicated multiband  structure with all five Fe $d-$bands crossing the Fermi level.
It has been proposed that  differences  in the $p-d$ hybridization may lead to the formation of more localized orbitals. \cite{wu} 
Therefore, each band could be affected by electron correlations differently to a degree that an orbital-selective  Mott transition may take place.\cite{Cap}  

In order to address the problem of  electronic correlations and possible charge localization, we focus on the \FeTe compound, a member of 
the layered iron-chalcogenide, Fe$Q$ ($Q$ = Se, Te) superconductors. The two end members, Fe$_{1.01}$Se and Fe$_{1+\delta}$Te ($\delta \leq 0.14$), exhibit 
fundamentally different ambient-pressure ground states. Fe$_{1+\delta}$Se is a superconductor with  critical temperature $T_c \sim 9$~K at 
ambient pressure.\cite{ref1,ref4,ref2} On the other hand, AF long-range order develops in Fe$_{1+\delta}$Te below $\sim 65$~K.\cite{ref8}  
The magnetic order vector $Q_{AF}=(\frac{1}{2},\frac{1}{2})$, the rather large ordered moment exceeding $2\, \mu_B$/Fe  and the Curie-Weiss-like susceptibility in the paramagnetic state of 
Fe$_{1+\delta}$Te~ \cite{sli} suggest that the magnetism is of a local-moment origin in contrast to the SDW phase found in other Fe-based superconductors. 
Moreover, inelastic neutron scattering measurements indicate a spin fluctuation spectrum, which is best described with 
an identical model to that used for the normal-state spin excitations in the high-$T_c$ cuprates.\cite{Lumsden}
In addition, an anomalously large mass renormalization, $m^*/m_{band}\approx 6-20$ has been reported recently for \FeTe from ARPES data,\cite{Tamai} consistent with the high bulk specific heat coefficient, $\gamma=39\, {\rm mJ/mol\, K^2}$.\cite{ref12} These results  highlight the importance of electronic correlations in the Fe$Q$ family, which  in analogy to other strongly correlated multiband systems may dramatically lower the energy difference between the coherent quasiparticle states and the incoherent excitations with more local character.\cite{Tamai}  

Here we report a combined EPR and $^{77}$Se,\Te NMR study of the FeSe${}_{0.42}$Te${}_{0.58}$ superconductor  ($T_c=11.5(1)\, {\rm K}$), which provide evidence for the coexistence of two electronic components 
arising from itinerant and localized states.  
The coupling between these states  at the atomic scale leads to the screening of localized moments, suppression of the AF spin fluctuations and thus opens the possibility for the emergence of unconventional superconductivity.
The intrinsic localized states are likely signatures of strong electron correlations  making the Fe$Q$ family a close relative to other high-$T_c$ superconductors.  

\begin{figure}[t]
\includegraphics[trim = 0mm 7mm 0mm 0mm, width=86mm,keepaspectratio=true]{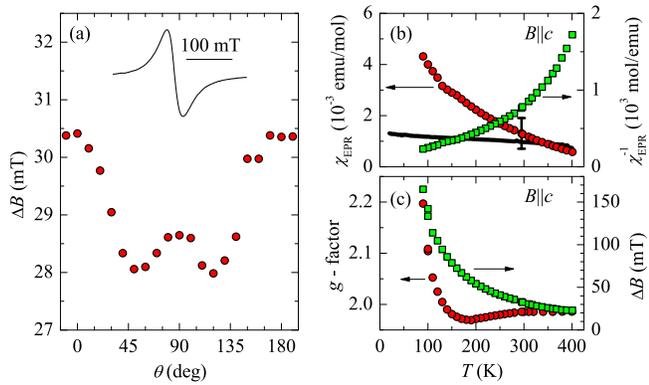}
\caption {(a) Angular dependence of the room temperature EPR linewidth in \FeTe single crystal. $\theta= 0$ is the $B||c$ crystal 
orientation. Inset: Room temperature EPR spectrum for $B||c$. Horizontal bar indicates field scale. (b) Temperature dependence of the 
EPR spin susceptibility, $\chi_{EPR}$ (circles, left scale), the bulk spin suscpetibility, $\chi_S$ (black solid line) and the inverse spin susceptibility, $\chi_{EPR}^{-1}$  (squares, right scale). 
(c) Temperature dependence of the EPR $g$-factor (left scale, circles) and  linewidth, $\Delta B$ (right scale, squares).}
\label{fig1}
\end{figure}

The single-crystalline \FeTe sample used in this work was identical to that of ref. \onlinecite{Tamai}. 
Hexagonal FeSe (1.27(2)\%) and elemental Se (2.31(4)\%)
were identified as impurities in crushed powders by synchrotron XRD measurements.
 The bulk magnetic susceptibility, $\chi_S$, was measured with a commercial Quantum Design MPMS system on a 0.547 mg single crystal and with the magnetic field  applied along the crystal $c-$axis. $\chi_S$ was determined by taking the difference between measurements performed in 3 and 2 T in order to subtract the contribution from ferromagnetic impurities. NMR frequency-swept spectra were measured in a magnetic field of 9.4~T  with a two-pulse sequence ${\beta}-{\tau}-{2\beta}-{\tau}-{\rm echo}$, a pulse length $\tau_{\beta}=4~\rm \mu$s and interpulse delay $\tau = 60~{\rm\mu}$s. For the temperature dependent X-band (9.6 GHz) cw-EPR experiments, a small piece of the crystal was exfoliated from the large crystal and sealed under dynamic vacuum in a 4-mm-diameter silica tube.

The very intense EPR resonance (inset Fig. \ref{fig1}a) has been measured at room temperature and is best described by Dyson lineshape as expected for metallic samples.  At 300 K, the calibrated  EPR intensity corresponds to a spin susceptibility, $\chi_{EPR} = 1.3(5)\times 10^{-3}$~emu/mol - the large uncertainty in the  value of $\chi_{EPR}$ arises from  difficulties in the precise positioning of the tiny single crystal in the resonator - which is comparable to the measured  $\chi_S = 1.0(1)\times 10^{-3}$~emu/mol (Fig. \ref{fig1}b) and to that reported for FeTe$_{0.55}$Se$_{0.45}$.\cite{ref12} The negligibly small refined content of Fe interstitials between the Fe(Se/Te) slabs\cite{Tamai} cannot be responsible for the measured $\chi_{EPR}$. On the other hand, hexagonal FeSe$_{1-x}$ phases can be ferromagnetic with Curie temperatures exceeding room temperature  \cite{hFeSe} and could give rise to strong ferromagnetic resonance. However, since the hexagonal FeSe$_{1-x}$ magnetization is already fully saturated in the field of the EPR resonance ($\sim 0.33$ T), it cannot account for the strong temperature dependence of  $\chi_{EPR}$ (Fig. \ref{fig1}b). We thus  conclude that the measured EPR signal is {\em intrinsic}.  

The EPR resonance shows a strong angular dependence of the EPR linewidth, $\Delta B$,  with a minimum value at 
the angle $\theta_m=54^o$ when the crystal is rotated away from the $B||c$-orientation (Fig. \ref{fig1}a). 
The minimum in  $\Delta B$ can be reproduced with the dipolar interactions between the exchange coupled localized moments centered at 
Fe-positions in the  FeSe$_{0.42}$Te$_{0.58}$ structure.  This implies that states with more local character may also exist in addition to the  quasiparticle states.   
 These are further evidenced by the temperature dependence 
of $\chi_{EPR}$, which rapidly increases with decreasing temperature (Fig. \ref{fig1}b). 
However, the non-linear dependence of the inverse EPR susceptibility  between 100 and 400 K (Fig. 1b) is not consistent with the simple 
Curie-Weiss law expected for localized moments only, meaning that the measured EPR signal has contributions from both quasiparticle and localised states.
Simple macroscopic phase segregation into metallic and insulating fractions would have implied 
that  $\chi_{EPR}$  can be expressed as $\chi_{EPR} = \chi_c + \chi_l$, where $\chi_c$ is the spin susceptibility of quasiparticles, which is expected to be only weakly temperature dependent and $\chi_l$ is the spin susceptibility of localized  states.  
But this approach results in unphysical parameters (negative $\chi_c$), thus leading to the conclusion that both electronic components  not only coexist at the nanometric or atomic scale but that they are also strongly coupled. Such coupling could be responsible for the rapid increase in $\Delta B$ and $g-$factor 
with decreasing temperature (Fig. \ref{fig1}c), which indicates development of local magnetic fields sensed by these states. It could also account for another surprising observation: namely, $\chi_{EPR}$($T$) is larger than the weakly temperature- dependent $\chi_S$ below room temperature  (Fig. \ref{fig1}b). If the coupling is strong enough, then localized states polarize conduction electrons and reduce the effective moment measured in bulk experiments.

\begin{figure}[t]
\includegraphics[trim = 0mm 10mm 0mm 5mm, width=86mm,width=86mm,keepaspectratio=true]{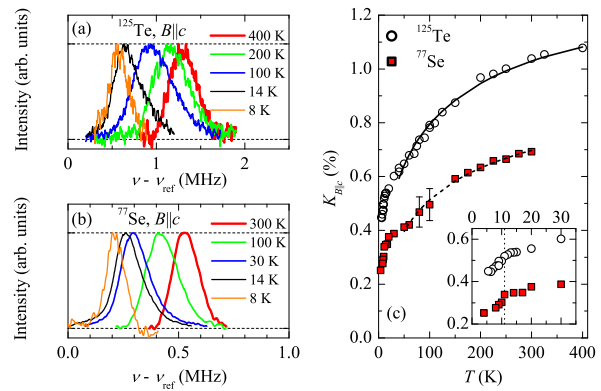}
\caption {Frequency-swept (a) \Te and (b) \Se NMR spectra of \FeTe single crystal measured with $B||c$. (c) Temperature dependence of the \Te ($^{125}K$, open circles) and \Se ($^{77}K$, solid squares) Knight shifts. 
The lines are fits to the model described in the text. Inset: expanded region near $T_c = 11.5(1)$~K, showing the clear drop of $^{125}K$ and $^{77}K$. }
\label{fig2}
\end{figure}

To confirm these hypotheses, we employed the NMR local probe technique, which can provide insight on the coexisting electronic components at different scales.  In addition,  it is 
not sensitive to impurities in the $\sim 1$\% range.
Figs. \ref{fig2}a,b show the \Te and \Se NMR spectra recorded for $B||c$.  The room temperature linewidths  of \Te and \Se resonances, 
$\delta^{125}\nu_{1/2}\approx 420$~kHz and $\delta^{77}\nu_{1/2}\approx 130$~kHz, respectively, imply that the local-site structural inhomogeneities resulting from the statistical Se/Te site occupation slightly broaden the \Te and \Se NMR spectra, e.g. with respect to the \Se NMR linewidth measured for Fe$_{1.01}$Se.\cite{ref5} 
For comparison, $\delta^{77}\nu_{1/2}$ is similar to those in Fe$_{1.04}$Se$_{0.33}$Te$_{0.67}$ \cite{ref14} or FeSe$_{0.92}$.\cite{ref15}  
Interestingly, the ratio, $\delta^{125}\nu_{1/2}/\delta^{77}\nu_{1/2}\approx 3.2$ is significantly larger than that of the corresponding gyromagnetic ratios, $^{125}\gamma / ^{77}\gamma =1.65$, which is consistent with differences in the $p-d$ hybridization between FeTe and FeSe.\cite{pd_hyb} 

\begin{figure}[t]
\includegraphics[trim = 0mm 6mm 0mm 0mm, width=86mm,keepaspectratio=true]{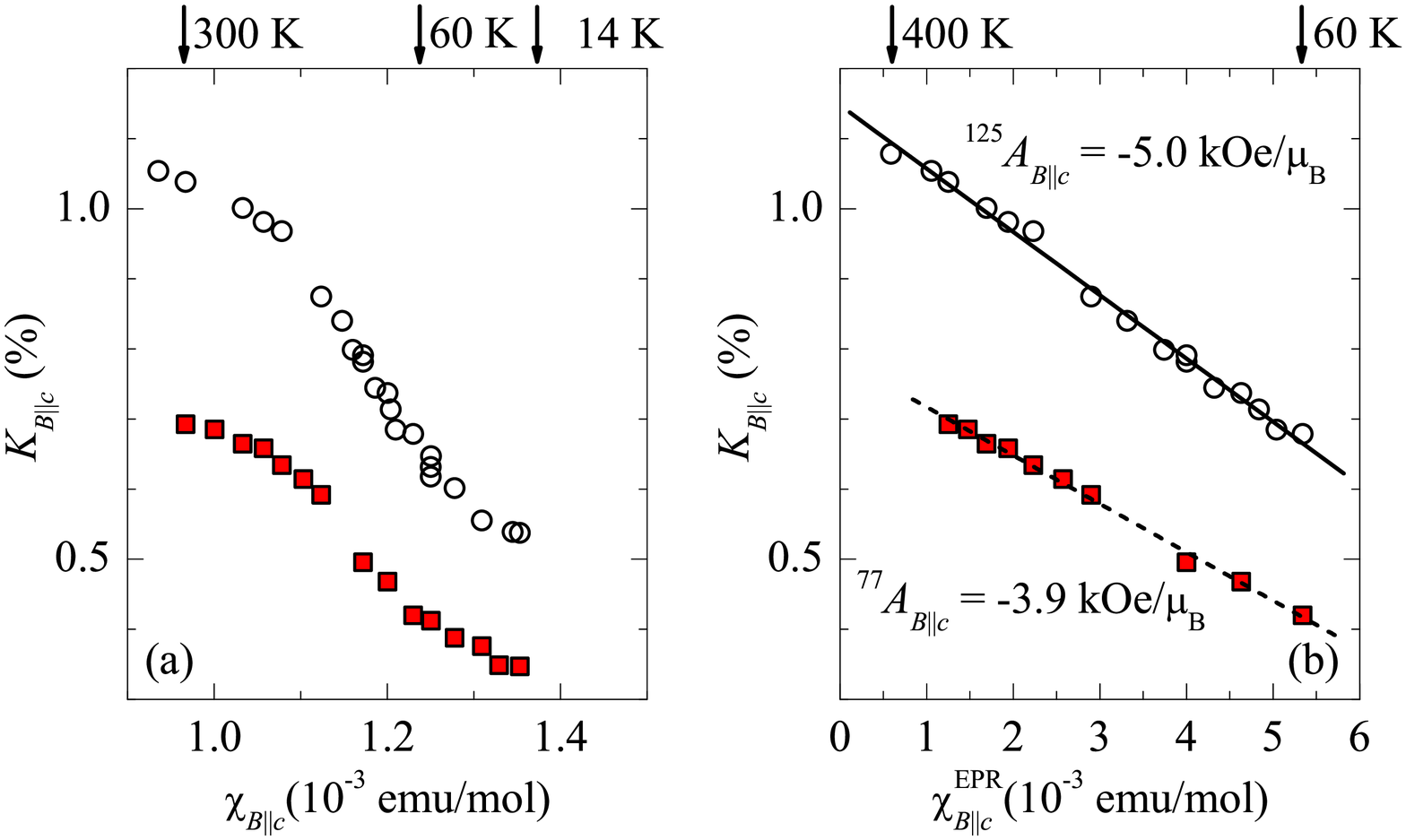}
\caption {$^{125}K$ and $^{77}K$ Knight shifts versus (a) bulk susceptibility, $\chi_{B||c}$ and (b) $\chi_{EPR}$ with temperature as an implicit parameter. 
 }
\label{fig3}
\end{figure}

The room temperature NMR spectra are strongly shifted to higher frequencies with respect to the reference. The  Knight shifts 
are $^{125}K=1.038(2)$\% and $^{77}K=0.692(2)$\% for \Te and \Se nuclei, respectively. 
The resonances shift considerably to lower frequencies with decreasing temperature (Fig. \ref{fig2}c): $\Delta^{125}K = -0.524$\% and  
$\Delta^{77}K = -0.316$\% between 300 and 20 K. A similar decrease of  $^{77}K$ has been reported for Fe$_{1.01}$Se and Fe$_{1.04}$Se$_{0.33}$Te$_{0.67}$.\cite{ref5,ref14,ref15}
However, the Knight shifts, $^nK$ ($n=77,125$) do not scale with the bulk spin susceptibility over the entire temperature range (Fig. \ref{fig3}a).  NMR data thus provide direct evidence that the local and bulk spin susceptibilities are different in the investigated sample. On the other hand, comparing
$^nK$ with $\chi_{EPR}$, which also  measures the local spin susceptibility, 
reveals excellent linear scaling  (Fig. \ref{fig3}b).  
If we express the temperature-dependent spin part of the Knight shift as $^nK (T)=\frac{^nA_{B||c}}{N_A\mu_B}\chi_{EPR}$,  
we derive the coupling constants, $^{125}A_{B||c}= -5.0(5)$~kOe/$\mu_B$ and $^{77}A_{B||c}=  -3.9(8)$~kOe/$\mu_B$. Since we scale $^nK$ 
with the local rather than the bulk spin susceptibility, the coupling constants are different from those extracted for 
Fe$_{1.04}$Se$_{0.33}$Te$_{0.67}$ only from low temperature ($<100$~K) data.\cite{ref14} 

The scaling of $^nK$  with the local rather than with the bulk spin susceptibility is a strong indication for the coexistence of coupled localized and itinerant states at the atomic scale.  
 In the case of two coupled spin components, there are generally three contributions to the spin part of the Knight shift, 
$^nK_S=^nK_c+^nK_l+^nK_{ex}$. Here $^nK_c$ stands for the coupling of Te/Se nuclei to the itinerant electrons via hyperfine 
coupling interaction and should be only weakly temperature dependent, $^nK_l$ describes the interaction with the localized states, and $^nK_{ex}$ is the additional Knight shift arising 
from the spin-density polarization due to the interaction between the localized  and  itinerant states.  
$^nK_{ex}$ should be negative in sign,\cite{ref20} as it is indeed observed.
It is  intriguing that the strong temperature dependence of $^nK$ can be  simulated with the expression
$
^nK\propto \left( 1- (T/T^*)\right)\log\left( T^*/T\right), 
$
 which has been applied to a number of Kondo lattice materials.\cite{ref21} Here $T^*$ is the correlated Kondo temperature and  is a measure of the intersite localized state interactions.\cite{ref21} Excellent agreement with the experimental data for both nuclei (Fig. \ref{fig2}b) is obtained with the same $T^* \sim 800$~K, which falls within the 50-80 meV range of the crossover energy between quasiparticle states and excitations with local character.\cite{Tamai} 

\begin{figure}[t]
\includegraphics[trim = 0mm 8mm 0mm 2mm, width=86mm,keepaspectratio=true]{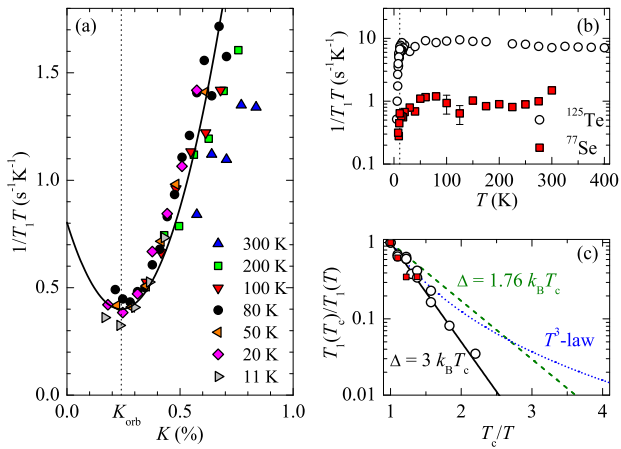}
\caption {(a)  Frequency dependence of $1/^{77}T_1T$ measured at various temperatures (bottom). All measurements fall on the same curve, which scales as $(^{77}K-^{77}K_{orb})^{2}$ with $^{77}K_{orb} = 0.24(3)$\% (solid line). 
(b) Temperature dependence of \Te (open circles) and \Se (solid squares) $1/^nT_1T$ rates. (c)  Temperature 
dependence of $T_1(T_c)/T_1(T)$ below $T_c = 11.5(1)$~K. The rate of suppression of $1/T_1$ below $T_c$ is significantly larger than expected for BCS-type superconductivity.}
\label{fig4}
\end{figure}

The most important experimental finding of this work is that in \FeTe intrinsic  states with  localized character may form, coexist and couple with itinerant states. The  coupling between the two electronic components governs the normal as well as the superconducting 
properties.  It is suggested that in such strongly correlated systems  this coupling plays a vital role in suppressing magnetism and 
promoting high-temperature superconductivity.\cite{ref22} Therefore, in order to test the suppression of spin fluctuations we turn 
to the spin-lattice relaxation time, $^nT_1$ data.  Fig. \ref{fig4}a shows the frequency dependence of $1/^{77}T_1T$ 
for \Se NMR spectra at selected temperatures. It is evident that $1/^{77}T_1T$ substantially varies over the \Se NMR line and 
the ratio between  largest and shortest $1/^{77}T_1T$ measured for the low- and high-frequency spectral 
shoulders can be as large as 4 (at 50 K). Therefore, a simple two relaxation-times model earlier applied\cite{ref14,ref15} to FeSe$_{0.92}$ and 
Fe$_{1.04}$Se$_{0.33}$Te$_{0.67}$ oversimplifies the experimental situation and may even lead to erroneous conclusions. 
All $1/^{77}T_1T$ values fall on nearly the same universal Knight shift-dependent curve described by the Korringa relation,  
$^{77}T_1T^{77}K_S^2=\frac{\hbar}{4\pi k_B}\frac{\gamma_e^2}{\gamma_{{\textrm Se}}^2}\beta$.  Here $\gamma_e$ and $\gamma_{{\textrm Se}}$ are the electron and nuclear gyromagnetic ratios and $^{77}K_S$ is obtained from $^{77}K_S=^{77}K-^{77}K_{orb}$, where $^{77}K_{orb} = 0.24(3)$\% is a temperature-independent  orbital shift.  Deviations from this universal dependence, mostly visible at high temperatures, indicate the presence of another weaker relaxation channel, for which $^{77}T_1^{-1}=const.$ might suggest relaxation via localized moments. A phenomenological parameter, $\beta$ characterizes the extent of spin fluctuations and has been recently studied in the context of the normal state properties of iron-based superconductors.\cite{LiFeAsRC} Since we obtain $\beta = 1.0(1)$, we conclude 
that AF spin fluctuations are not strongly enhanced.  Another indication for this comes from the validity of Korringa relation over a broad temperature  interval (Fig. \ref{fig4}b).  
This is in striking contrast to FeSe, which clearly shows enhancement of AF spin fluctuations 
towards $T_c$.\cite{ref5} Apparently these spin fluctuations become strongly suppressed upon Te substitution and are only visible 
again after the application of pressure.\cite{shim} This is fully consistent with the present picture of localized states. 
Namely high $T^*$ suggests strong local Kondo effects where local magnetic moments become screened and unconventional 
superconductivity may develop instead of magnetism. 

Last, we  turn to the \Te and \Se NMR data below $T_c=11.5(1)$~K. The \Te resonance suddenly becomes narrower and more 
symmetric, while its intensity starts to decrease (Fig. \ref{fig2}a). The abrupt decrease in the signal intensity is due to the Meissner 
shielding of the $rf$ pulses. On the other hand, the sudden decrease in the linewidth is more surprising.  
In the singlet superconducting state, we expect $\chi_S$  to vanish and therefore any broadening and extra resonance shift caused by 
the interaction between the localized and superconducting states should be reduced below $T_c$.  $^{125}K$ and $^{77}K$ suddenly start to decrease at faster rate below $T_c$ (inset Fig. \ref{fig2}b), thus  indicating  the vanishing spin susceptibility 
as expected both for $s-$ and $d-$wave pairing.  This is further supported by  the $^nT_1$ data. 
$^nT_1^{-1}$ values are strongly reduced below $T_c$ for both nuclei (Fig. \ref{fig4}c). We also note that $1/T_1$ does not show a 
coherence peak which has been also missing in FeSe \cite{ref5} and other Fe-based superconductors.\cite{ref15, ref26, ref27, ref28}
$1/^nT_1$  is exponentially suppressed below $T_c$ with an effective gap, $\Delta = 3 k_BT_c$ which is much larger than that predicted for conventional $s-$wave BCS-type  pairing in the weak-coupling limit, $\Delta = 1.76 k_BT_c$. 
Neither the low-temperature  $T^3$ dependence (expected within the $s_\pm$ gap scenario) nor the two-gap dependence is found down to $T_c/T=2.25$, in striking contrast to reports for Fe-pnictide superconductors.
We also stress that the obtained $\Delta$ is in excellent agreement with that found by point-contact Andreev reflection spectroscopy ($\Delta = 3.1 k_BT_c$) \cite{Andr} and implies  a single $s-$wave order parameter in the strong coupling limit. An alternative explanation would be in terms of considerable anisotropy of the superconducting gap frequently noticed in systems with strong electronic correlations.\cite{schl} Additional experiments are needed to verify this scenario.   

The detection of intrinsic localized moments  coupled to itinerant electrons and leading to unconventional SC shows some similarities 
with strongly correlated electron systems.  The question to resolve is, how these localized states form in FeSe$_{0.42}$Te$_{0.58}$. Orbitally selective Mott 
localization has been recently proposed \cite{Cap} for Fe-based superconductors. This model  could well explain the strong local moment 
screening and suppression of AF fluctuations in the normal state as well as the  \Te and \Se NMR data below $T_c$.  The coexisting magnetic 
and superconducting order parameters on the atomic scale that have been recently suggested  for FeSe  from $\mu$SR experiments \cite{ref7} 
is also consistent with this picture. 
All these results point to the importance of strong intraband electronic correlations which may explain the rapid suppression of 
$1/T_1$  below $T_c$, the  strong sensitivity of  FeSe$_{1-x}$Te$_x$ superconductivity both to chemical substitution and applied pressure \cite{serPRB,natmat,gresty,mizu} and the induced static magnetic order 
at pressures exceeding 1 GPa.\cite{ref7}   

In conclusion, we have carried out EPR and NMR studies of \FeTe single crystal. We found the presence of intrinsic localized states 
coupled to quasiparticles. The screening of these  states suppresses the AF spin fluctuations and thus opens the 
possibility for the emergence of unconventional superconductivity. Although the exact origin of localized states should be 
investigated in the future, the present picture is consistent with  the intraband electronic correlations leading to a localization 
of one of the Fe-derived bands. In this respect, the Fe$Q$ family appears to be much closer to other high-$T_c$ superconductors and 
should be treated on a similar footing.


\begin{thebibliography}{99}
\bibitem{cup} P.A. Lee et al., {\em Rev. Mod. Phys.} {\bf 78}, 17 (2006).
\bibitem{sci} Y. Takabayashi et al., {\em Science} {\bf 323}, 1585 (2009).
\bibitem{nat} A. Y. Ganin et al., {\em Nature}, doi:10.1038/nature09120. 
\bibitem{yang} W.L. Yang et al., {\em Phys. Rev. B} {\bf 80}, 014508 (2009).
\bibitem{wu} J. Wu et al., {\em Phys. Rev. Lett.} {\bf 101}, 126401 (2008).
\bibitem{Cap} L. de' Medici et al., {\em J. Supercond. Novel Magn.} {\bf 22}, 535 (2009) .
\bibitem{ref1} F.C. Hsu et al.,  {\em Proc. Natl. Acad. Sci.} {\bf 105}, 14262 (2008).
\bibitem{ref4} S. Margadonna et al., {\em Chem. Commun.} 5607 (2008).
\bibitem{ref2} T.M. McQueen et al., {\em Phys. Rev. B} {\bf 79}, 014522 (2009).
\bibitem{ref8} Y. Xia et al., {\em Phys. Rev. Lett.} {\bf 103}, 037002 (2009).
\bibitem{sli} S. Li et al., {\em Phys. Rev. B} {\bf 79}, 054503 (2009).
\bibitem{Lumsden} M.D. Lumsden et al., {\em Nat. Phys.} {\bf 6}, 182 (2010).
\bibitem{Tamai} A. Tamai et al., {\em Phys. Rev. Lett.} {\bf 104}, 097002 (2010).
\bibitem{ref12} B. C. Sales, et al. {\em Phys. Rev. B} {\bf 79}, 094521 (2009).
\bibitem{hFeSe} T. Kamimura and T. Iwata, {\em J. Phys. Soc. Jpn.} {\bf 45} 1769 (1978).
\bibitem{ref5} T. Imai et al., {\em Phys. Rev. Lett.} {\bf 102}, 177005 (2009).
\bibitem{ref14} C. Michioka et al., arXiv:0911.3729.
\bibitem{ref15} H. Kotegawa et al., {\em J. Phys. Soc. Jpn.} {\bf 77}, 113703 (2008).
\bibitem{LiFeAsRC} P. Jegli\v c et al., {\em Phys. Rev. B} {\bf 81}, 140511(R) (2010).
\bibitem{pd_hyb} T. Miyake et al., {\em J. Phys. Soc. Jpn.} {\bf 79}, 044705 (2010).
\bibitem{ref20} J.B. Boyce and C.P. Slichter, {\em Phys. Rev. B} {\bf 13}, 379 (1976).
\bibitem{ref21} N.J. Curro et al., {\em Phys. Rev. B} {\bf 70}, 235117 (2004).
\bibitem{ref22} F. J. Ohkawa, {\em J. Phys. Soc. Jpn} {\bf 61}, 4490 (1992).
\bibitem{shim} Y. Shimizu et al., {\em J. Phys. Soc. Jpn.} {\bf 78}, 123709 (2009).
\bibitem{ref26}   H.-J. Grafe et al., {\em New. J. Phys.} {\bf 11}, 035002 (2009).
\bibitem{ref27}   F.L. Ning et al., {\em J. Phys. Soc. Jpn} {\bf 77}, 103705 (2008).
\bibitem{ref28}   Y. Nakai et al., {\em J. Phys. Soc. Jpn} {\bf 77}, 073701(2008).
\bibitem{Andr} W.K. Park et al., arXiv:1005.0190.
\bibitem{schl} D.F. Smith and C.P. Slichter, {\em Lect. Notes Phys.} {\bf 684}, 243 (2006).
\bibitem{ref7} M. Bendele et al., {\em Phys. Rev. Lett.} {\bf 104}, 087003 (2010).
\bibitem{serPRB} S. Margadonna et al., {\em  Phys. Rev. B} {\bf 80}, 064506 (2009).
\bibitem{natmat} S. Medvedev et al., {\em Nature Mater.} {\bf 8}, 630 (2009).
\bibitem{mizu} Y. Mizuguchi et al., {\em Appl. Phys. Lett.} {\bf 94}, 012503 (2009).
\bibitem{gresty} N.C. Gresty et al., {\em J. Am. Chem. Soc.} {\bf 131}, 16944 (2009).



\end{thebibliography}
\end{document}